\documentclass{article}
\pdfoutput=1
\usepackage{spconf,amsmath,graphicx}
\usepackage{cite}
\usepackage{amssymb,amsfonts}
\usepackage{algorithmic}
\usepackage{textcomp}
\usepackage{xcolor}
\usepackage[acronym]{glossaries}

\usepackage{makecell}
\usepackage{cancel}
\usepackage{amsthm}
\usepackage{textcomp}
\usepackage{booktabs,ragged2e}
\usepackage{multirow}
\usepackage{bm}
\usepackage[normalem]{ulem}
\usepackage{siunitx}

\newacronym{gnss}{GNSS}{Global Navigation Satellite System}
\newacronym{gdop}{GDOP}{Geometric Dilution of Precision}
\newacronym{gps}{GPS}{Global Positioning System}
\newacronym{rss}{RSS}{received signal strength}
\newacronym{crb}{CRB}{Cramér-Rao bound}
\newacronym{mle}{MLE}{maximum likelihood estimator}
\newacronym{nn}{NN}{neural network}
\newacronym{mse}{MSE}{mean squared error}
\newacronym{kf}{KF}{Kalman filter}
\newacronym{apbm}{APBM}{augmented physics-based model}
\newacronym{pl}{PL}{pathloss}
\newacronym{inr}{INR}{interference-to-noise ratio}
\newacronym{rmse}{RMSE}{root-mean-square error}

\usepackage{color, colortbl}
\definecolor{Gray}{gray}{0.9}
\definecolor{LightCyan}{rgb}{0.88,1,1}
\newcolumntype{g}{>{\columncolor{Gray}}c}

\theoremstyle{definition}

\newcommand{\Pn}{y_n}
\newcommand{\Pvec}{\mathbf{y}}
\newcommand{\Ptx}{P_0}
\newcommand{\Pz}{P_0}

\newcommand{\xn}{\mathbf{x}_n}
\newcommand{\xnel}[1]{{x}_n^{(#1)}}
\newcommand{\wn}{\xi_n}
\newcommand{\ccov}{\mathbf{C}}
\newcommand{\param}{\bm{\theta}}
\newcommand{\nnpar}{\bm{\phi}}
\newcommand{\paramel}[1]{\theta_{#1}}
\newcommand{\paramelhat}[1]{\hat{\theta}_{#1}}
\newcommand{\dist}{d(\xn,\param)}
\newcommand{\distSq}{d^2(\xn,\param)}
\newcommand{\distFo}{d^4(\xn,\param)}
\newcommand{\yvec}{\mathbf{y}}

\newcommand{\inr}{\text{INR}}
\newcommand{\rmse}{\text{RMSE}}
\newcommand{\nearfield}{d_F}

\newcommand{\fish}{\bm{I}}
\newcommand{\expec}{E}
\newcommand{\trsp}{^{\top}}
\newcommand{\eye}{I}

\DeclareMathOperator{\tr}{tr}

\DeclareMathOperator{\var}{var}

\title{JAMMING SOURCE LOCALIZATION USING AUGMENTED PHYSICS-BASED MODEL}
\name{Andrea Nardin$^{\ast}$, Tales Imbiriba$^{\dagger}$, Pau Closas$^{\dagger}$\thanks{This work has been partially supported by the NSF under Award ECCS-1845833.}}
\address{$^{\ast}$ Dept. of Electronics and Telecommunications, Politecnico di Torino, Turin, Italy \\
$^{\dagger}$ Dept. of Electrical \& Computer Engineering, Northeastern University, Boston, MA (USA)}

\begin{document}
\ninept
\maketitle
\begin{abstract}
Monitoring interferences to satellite-based navigation systems is of paramount importance in order to reliably operate critical infrastructures, navigation systems, and a variety of applications relying on satellite-based positioning. This paper investigates the use of crowdsourced data to achieve such detection and monitoring at a central node that receives the data from an arbitrary number of agents in an area of interest. Under ideal conditions, the pathloss model is used to compute the Cramér-Rao Bound of accuracy as well as the corresponding maximum likelihood estimator. However, in real scenarios where obstructions and reflections are common, the signal propagation is far more complex than the pathloss model can explain. We propose to augment the pathloss model with a data-driven component, able to explain the complexities of the propagation channel. The paper shows a general methodology to jointly estimate the interference location and the parameters of the augmented model, showing superior performances in complex scenarios such as those encountered in urban environments.
\end{abstract}
\begin{keywords}
Jamming localization, GNSS, augmented physics-based model, neural networks. 
\end{keywords}
\section{Introduction}
\label{sec:intro}

GNSS is a general term that encompasses satellite-based navigation systems such as GPS, Galileo, Glonass and Beidou among others \cite{DarCloDju:J15,morton2021position}. It relies on a constellation of satellites synchronously emitting known signals, which enable computation of a receiver position, velocity, and time (PVT) unknowns. GNSS is arguably the primary technology when it comes to position, navigation, and timing (PNT) applications, such as intelligent transportation systems \cite{Yastrebova2021,williams2022impact,minetto2019collab}, critical infrastructures \cite{ioannides2016known}, environmental applications
\cite{Imam2019}, or agriculture applications \cite{Wang2019} to name a few.
However, the increasing dependence on GNSS has created a concern about its potential vulnerabilities \cite{amin2016vulnerabilities}, mostly related to GNSS vulnerability to simple jamming interferences \cite{borio2016impact,Ferre2020} or more complex spoofing attacks \cite{sathaye2020semperfi}. 

It is therefore of crucial importance to develop interference monitoring systems that can detect and localize the possible threats in an area \cite{dovis2015gnss,thombre2018gnss}. 
In this paper we are particularly interested in congested areas, where malicious users are more prone to operate, where crowdsourced \cite{arias2018crowd} data can be leveraged in order to implement such monitoring. In particular, we foresee a system where agents navigate an area with capabilities of transmitting the measured signal power at the corresponding GNSS frequency bands. Recall that GNSS signals are transmitted using a spread-spectrum system such that they are below the receiver's noise floor, as a consequence signal values with powers larger than the noise floor would likely correspond to a jamming signal being present in the vicinity. If a large number of agents convey this information to a central unit, the data can be processed in order to localize the source of jamming interferences and, ultimately, take corrective measures. Several challenges of this approach are addressed in this paper, namely the need for an accurate signal propagation model from which one can perform inference from. The nominal pathloss model has severe disadvantages when it comes to complex environments, given that accounting for reflections and obstructions is not simple. To overcome this, we propose to augment the pathloss model with a data-driven component (e.g. a neural network, or NN for short) that is able to learn the position dependent parts of the propagation model that are complex to accurately model. We refer to these models as \glspl{apbm}, which were succesfully used in other contexts \cite{Demirkaya_hybridCKF_2021,Li_DeepAECUnmixing2021,Imbiriba_hybrid_2022}.

Section \ref{sec:problem} presents the jammer localization problem, discussing the nominal pathloss signal propagation model that is used as a measurement equation in the proposed scheme. Under the nominal model, the \gls{crb} of estimation accuracy is derived as well as the \gls{mle}. The proposed APBM augmentation to the pathloss is discussed in Section \ref{sec:apbm}, where it is shown how we propose to jointly estimate the jammer's location and the parameters of the augmented model. Experiments are provided in Section \ref{sec:exp} for $i)$ an open-sky scenario where the pathloss is an accurate model, which is used to validate the proposed methods; and $ii)$ a more complex urban scenario, where the received signal powers are generated using ray tracing techniques, where the APBM shows its superiority to plain pathloss modeling. Finally, Section \ref{sec:conclusion} discusses the main conclusions and outlook of this work.

\section{Pathloss-based estimation of jammer's position}\label{sec:problem}


\subsection{The pathloss measurement model}
We propose a crowdsourced framework where the position of the jammer is estimated from a sequence of $N$ observations of the jammer power at several locations. In particular, the $n$-th observation $\Pn$ (dBW) is related to the vector of parameters we want to estimate through a generic function $f(\xn; \param)$, such that
\begin{equation}
\label{eq:general_model}
    \Pn = f(\xn; \param) + \wn
\end{equation}
where $\xn = (\xnel{1},\dots, \xnel{D})^\top$ is the location where the measurement took place, $\param = (\theta_1,\dots,\theta_D)^\top$ is the jammer location in $D$ dimensions (often 3), and $\wn$ is the measurement noise, which is assumed to be additive and independent of $\xn$ and $\param$. This measurement model is particularly relevant in the context of GNSS interference monitoring since useful signals are received below the noise floor and large signal powers can be regarded as jamming signals. 

A common model for \gls{rss} observations is the pathloss model~\cite{wu2019wifi}:
\begin{equation}
\label{eq:rss_model}
    f(\xn; \param) = \Pz - \gamma 10\log_{10} \dist
\end{equation}
where $\Pz$ (dBW) is the jammer power at the reference distance of 1\,m, and $\dist$ is the distance between the $n$-th observer at $\xn$ and the jammer~\cite{rappaport2002}. It is defined as
\begin{equation}
    \dist = \|\xn-\param\| = \sqrt{(\xn-\param)\trsp(\xn-\param) }
\end{equation}
with $\| \cdot \|$ being the Euclidean norm.

\subsection{The Cramér-Rao bound}
The \gls{crb} of the $i$-th parameter is defined through the Fisher information matrix (FIM) \cite{kay1993fundamentals} as
\begin{equation}
\label{eq:fisher}
    [\fish(\param)]_{ij} = - \expec \left[ \frac{\partial^2 \ln p{(\Pvec;\param)}}{\partial \theta_i \partial \theta_j} \right]
\end{equation}
where $p{(\Pvec;\param)}$ is the \emph{likelihood} function of the observed data and $\Pvec=(y_1,\dots,y_N)^\top$ is the vector of observations. As a result
\begin{equation}\label{eq:crb_def}
    \var([\hat{\param}]_i) \geq [\fish^{-1}(\param)]_{ii}
\end{equation}
gives the minimum variance attainable by an unbiased estimator $\hat{\param}$ of the parameters $\param$.

In~\eqref{eq:rss_model}, the unpredictable shadowing effects are modeled by $\wn$ and are experienced by measurements which have the same locations, but different clutter on the propagation path. It has been shown that in nominal conditions this measurement noise can be modeled with a log-normal distribution~\cite{bernhardt1987macroscopic,cox1984800mhz}, thus resulting in $ \wn \sim \mathcal{N}(0,\sigma) $ when values are expressed in dB as in~\eqref{eq:rss_model}.
%
As a result, the vector of observations $\Pvec$ is distributed according to
\begin{equation}
\label{eq:obs_normal}
    \Pvec | \param \sim \mathcal{N}(\bm{\mu}(\param),\ccov(\param))
\end{equation}
We can thus resort to a generalized \gls{crb} for the general Gaussian case~\cite{kay1993fundamentals}.
When~\eqref{eq:obs_normal} holds, equation~\eqref{eq:fisher} becomes
\begin{multline}
\label{eq:fisher_general_gaussian}
     [\fish(\param)]_{ij} = \left[ \frac{\partial \bm{\mu}(\param)}{\partial \theta_i} \right]\trsp \ccov^{-1}(\param) \left[ \frac{\partial \bm{\mu}(\param)}{\partial \theta_j} \right]\\ + \frac{1}{2} \tr \left[ \ccov^{-1}(\param) \frac{\partial \ccov(\param)}{\partial \theta_i} \ccov^{-1}(\param) \frac{\partial \ccov(\param)}{\partial \theta_j} \right]\ \;,
\end{multline}
\noindent which under the assumption of independent measurements and constant variance we get $\ccov(\param) = \sigma \eye$, can be simplified to
\begin{align}
     [\fish(\param)]_{ij} &= \left[ \frac{\partial \bm{\mu}(\param)}{\partial \theta_i} \right]\trsp \frac{1}{\sigma^2} \eye \left[ \frac{\partial \bm{\mu}(\param)}{\partial \theta_j} \right]\\
     &= \frac{1}{\sigma^2} \sum_{n=1}^{N}  \frac{\partial [\bm{\mu}(\param)]_n}{\partial \theta_i} \frac{\partial [\bm{\mu}(\param)]_n}{\partial \theta_j}\\
     &= \frac{1}{\sigma^2} \sum_{n=1}^{N}  \frac{\partial f(\xn;\param)}{\partial \theta_i} \frac{\partial f(\xn;\param)}{\partial \theta_j}\ .
\end{align}
Using the particular expression of the pathloss model in \eqref{eq:rss_model}, the FIM can be further particularized to
\begin{align}
    \frac{\partial f(\xn;\param)}{\partial \paramel{i}} &= -10 \gamma \frac{\partial \log_{10}(\dist)}{\partial \paramel{i}}\\
    &= \frac{-10 \gamma}{\dist \ln{10}} \frac{(2 \paramel{i}- 2 \xnel{i})}{2\dist}\\
    &= \frac{-10 \gamma}{\ln{10}} \frac{( \paramel{i}-\xnel{i})}{\distSq}\ .
\end{align}
The elements of the FIM then become
\begin{equation}
    [\fish(\param)]_{ij} = \frac{100 \gamma^2}{\sigma^2 (\ln{(10)})^2} \sum_{n=1}^{N}  \frac{(\paramel{i}-\xnel{i})(\paramel{j}-\xnel{j})}{\distFo}
\end{equation}
resulting in
\begin{equation}\label{eq:fisher_multidim}
    \fish(\param) = \frac{100 \gamma^2}{\sigma^2 (\ln{(10)})^2} \sum_{n=1}^{N} \frac{1}{\distFo}(\param - \xn)\trsp(\param - \xn) \ .
\end{equation}

The CRB for the $i$-th element of $\hat{\param}$ can be obtained by replacing~\eqref{eq:fisher_multidim} into~\eqref{eq:crb_def}.
For the 2-dimensional case we can derive a simple analytic CRB for the estimator of the jammer's position. Thus, the inverse of $\fish(\param)$ can be easily computed leading to 
\begin{equation}
    \var({\paramelhat{1}}) \geq \frac{\sigma^2(\ln{(10)})^2}{100 \gamma^2} \frac{b}{ab-c^2}
\end{equation}
\begin{equation}
    \var({\paramelhat{2}}) \geq \frac{\sigma^2(\ln{(10)})^2}{100 \gamma^2} \frac{a}{ab-c^2} \,\cdot
\end{equation}

where we defined
\begin{equation}
    a = \sum_{n=1}^{N}  \frac{\ (\paramel{1} - \xnel{1})^2}{\distFo}
\end{equation}
\begin{equation}
    b = \sum_{n=1}^{N}  \frac{\ (\paramel{2} - \xnel{2})^2}{\distFo}
\end{equation}
\begin{equation}
    c = \sum_{n=1}^{N}  \frac{\ (\paramel{1} - \xnel{1})(\paramel{2} - \xnel{2})}{\distFo}\ .
\end{equation}


\subsection{Maximum Likelihood Estimator}\label{sec:mle}
In this section we present the MLE estimator for the jammer's position $\param$, considering an arbitrary RSS function $f(\xn; \param)$.
For this, let us define $\mathbf{X}=\{\mathbf{x}_1,\dots,\mathbf{x}_N\}$ and $\mathcal{D}=\{y_n,\xn\}_{n=1}^{N}$ as a dataset composed of $N$ i.i.d. pairs of \textit{loci} $\xn$ and RSS measurements $y_n$.
%
Assuming~\eqref{eq:obs_normal} to define the statistical characteristics of the model in~\eqref{eq:general_model}, the likelihood distribution can be written as:
\begin{align}
\label{eq:lh_obsvec}
    p(\Pvec | &\mathbf{X},\param) = \prod_{n=1}^N p(\Pn | \xn,\param) \nonumber\\
    &= \frac{1}{(2\pi \sigma^2)^{N/2}} \exp{ \left\{ -\frac{1}{2\sigma^2} \sum_{n=1}^N (\Pn-f(\xn;\param))^2 \right\} }\ .
\end{align}
The log-likelihood function is therefore
\begin{equation}
\label{eq:llh}
    \ln{p(\Pvec | \mathbf{X},\param)}= -\frac{N}{2} \ln{(2\pi \sigma^2)} -\frac{1}{2\sigma^2} \sum_{n=1}^N (\Pn-f(\xn;\param))^2
\end{equation}
The estimator for the jammer's location, $\hat{\param}$, can be found by maximizing the $\log$-likelihood as
\begin{equation} \label{eq:maxloglik}
    \hat{\param}_\mathrm{MLE} = \mathop{\arg\max}_{\param} \ln p(\yvec | \mathbf{X},\param) \;.
\end{equation}



\begin{figure}[htb]
  \centering
  \includegraphics[width=0.47\columnwidth]{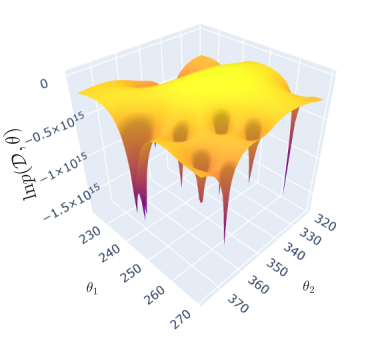}~
  \includegraphics[width=0.47\columnwidth]{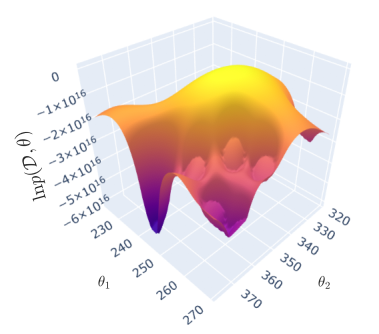}
 \vspace{-0.2cm}
\caption{Log-likelihood with respect to different $\param$, measured under an ideal pathloss propagation scenario with $f(\xn;\param)$ (\textbf{left panel}) and $\bar{f}(\xn;\param)$ (\textbf{right panel}).}
\label{fig:llh_singularities}
\end{figure}

One drawback of the pathloss model is that $f(\xn;\param)\to \infty$ when $\dist \to 0$. This generate singularities in the likelihood function that need to be addressed when solving the optimization in~\eqref{eq:maxloglik}. 
To circumvent this issue we modified~\eqref{eq:rss_model} to
\begin{equation}
\label{eq:rss_model_nearfield}
    \bar{f}(\xn; \param) = \Pz - \gamma 10\log_{10}\{ \max(\dist,\nearfield) \}
\end{equation}
where $\nearfield$ is the far-field distance, a limit that depends on the characteristics of the transmitting antenna and above which the far-field assumption that motivates the pathloss formulation in~\eqref{eq:maxloglik} is considered to hold~\cite{Rappaport96}.
The effect of this choice is twofold. First, infinite values are removed from the observation function and singularities are avoided. Secondly, the ``holes'' visible in Figure~\ref{fig:llh_singularities} (left panel) are filled before the gradient becomes too steep, towards the singularity points. In this way the maximization of the log-likelihood can be pursued efficiently through gradient-based methods. The resulting log-likelihood is shown in Figure~\ref{fig:llh_singularities} (right panel). This strategy is also effective when minimizing the cost function described next.


\vspace{-0.5cm}
\section{Augmented physics-based estimation of jammer's position}\label{sec:apbm}
Although the pathloss model in~\eqref{eq:general_model} is well suited to describe the jamming RSS field over open-spaces, it fails to capture the complexity of urban or indoor scenarios where multiple reflections and attenuations might occur. 
Additionally, our practical experience indicated that naive, in the sense that no prior information regarding the jamming field is incorporated in the model, NNs can only represent the jamming field accurately enough if humongous amounts of data is readily available. For this reason, in this section we propose an augmented physics-based (i.e., pathloss-based) neural network model where the NN acts as a nonlinear correction term around the pathloss function. 
Recently, APBMs have been used successfully in different scenarios~\cite{Santos_gp_crowdsourcing_tsp_2019, Demirkaya_hybridCKF_2021, Li_DeepAECUnmixing2021, Imbiriba_hybrid_2022}. 
Here, we propose to augment the modified pathloss model with NNs leading to an enhanced model:
\begin{equation}
\label{eq:augmented_model}
    h(\xn; \param, \nnpar) = \bar{f}(\xn; \param) + g(\xn; \nnpar)
\end{equation}
where $\nnpar\in\mathbb{R}^{M}$ is the vector of neural network's parameters.
The measurement model in~\eqref{eq:general_model} can then be re-written as 
\begin{equation}
    y_n = h(\xn; \param, \nnpar) + \wn \;,
\end{equation}
\noindent and used to perform inference as described later in this section.

To limit the NN from overpowering the physics-based model, the authors in~\cite{Li_DeepAECUnmixing2021, Imbiriba_hybrid_2022} considered different regularizations employed to force the NN to act as a correction term around the physics-based model, maintaining interpretability of the overall model. Similarly, we consider a $\ell_2$ regularization to control the NN contribution. 
Thus, 
we define the regularized cost function as 
\begin{equation}
    \mathcal{C}(\mathcal{D},\param,\nnpar) = \sum_{n=1}^N\|\yvec_n-h(\xn;\param,\nnpar)\|^2_2 + \beta \|\nnpar\|^2_2 
\end{equation}
where $\beta\in\mathbb{R}_+$ is a scalar controlling the regularization over the NN's parameters, such that when $\beta=0$ the NN dominates the APBM and when $\beta\rightarrow\infty$ the NN contribution is reduced.
Finally, we define the optimization problem with respect to the jammer location $\param$ and the NN's parameters $\nnpar$ as a empirical risk minimization problem as:
\begin{equation}\label{eq:apbm_inference}
    (\hat{\param}, \hat{\nnpar} ) = \mathop{\arg\min}_{\param, \nnpar} \mathbb{E}_{\hat{p}(\mathcal{D})}\bigg\{\mathcal{C}(\mathcal{D},\param,\nnpar)\bigg\}
\end{equation}
where the expectation operator is taken with respect to the empirical data distribution $\hat{p}(\mathcal{D})$.

\vspace{-0.25cm}
\section{Experiments}
\label{sec:exp}

In this section we present our experimental results in two simulated scenarios: an ideal scenario, where the jamming signal propagates according to a nominal \gls{pl} model, defined by $\gamma=2$ and described by \eqref{eq:rss_model}; and an urban scenario, subject to intense multipath and shadowing effects. 
The urban scenario is modeled exploiting a ray tracing approach~\cite{schaubach1992,yun2015}, computing multiple propagation paths using 3-D environment geometry and electromagnetic analysis, including free-space loss and reflection losses up to 4 reflections.

\vspace{-0.3cm}
\begin{figure}[htb]
  \centering
  \centerline{\includegraphics[width=0.45\textwidth]{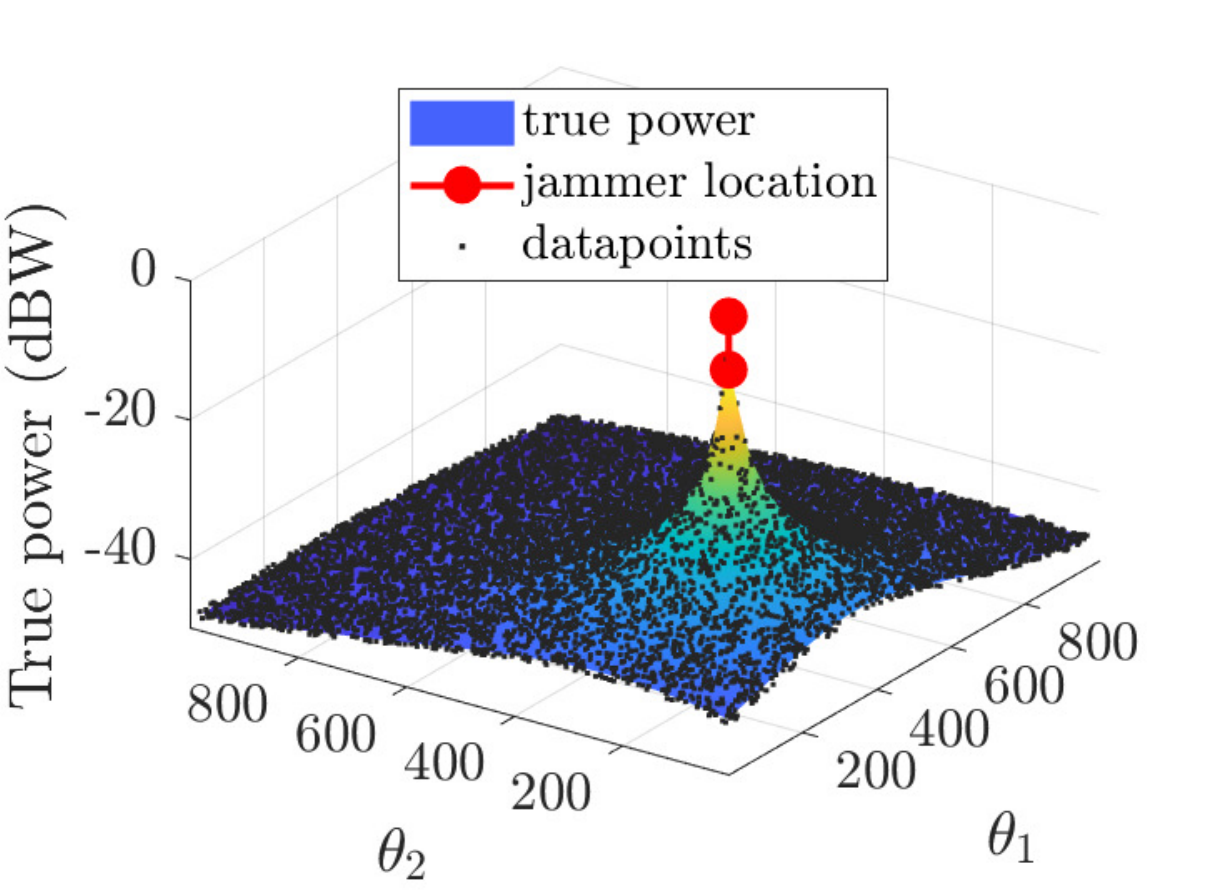}}
\vspace{-0.3cm}
\caption{Power measurements under nominal path loss propagation.}
\label{fig:PL_field}
\vspace{-0.8cm}
\end{figure}

\begin{figure}[htb]
  \centering
  \centerline{\includegraphics[width=0.45\textwidth]{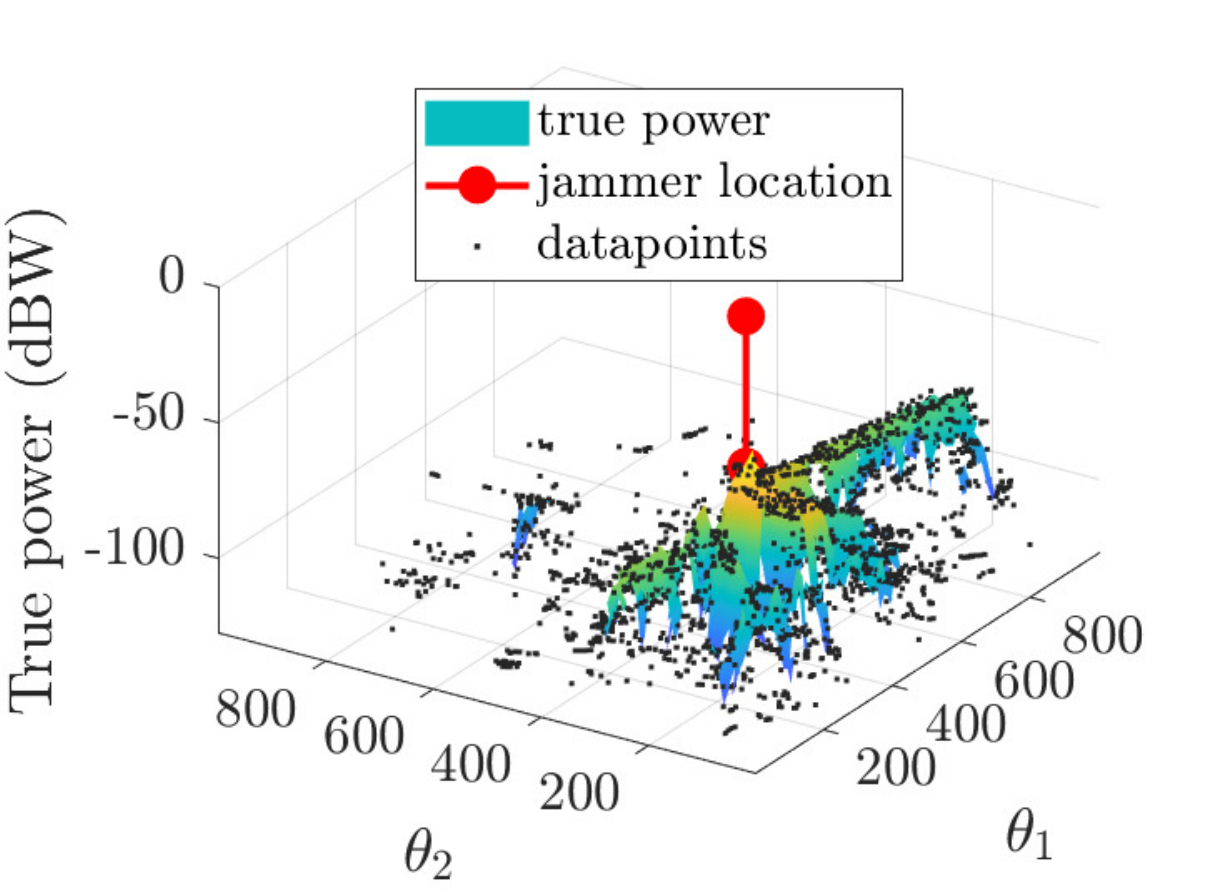}}
\vspace{-0.3cm}
\caption{Power measurements under ray tracing model propagation.}
\label{fig:RT_field}
%
\end{figure}

Particularly, we compare the localization performance of 
$(i)$ the MLE in \eqref{eq:maxloglik} under the PL model assumption; 
$(ii)$ the proposed APBM ($\beta=1$) implemented as in \eqref{eq:apbm_inference}; 
$(iii)$ the APBM ($\beta=1$) where the parameter $\Ptx$ is also estimated jointly along $\param$ and $\nnpar$, which is of practical importance;
$(iv)$ a PL-only solution for \eqref{eq:apbm_inference} where $h(\xn; \param, \nnpar) = \bar{f}(\xn; \param)$; and 
$(v)$ a NN-only solution for \eqref{eq:apbm_inference} where $h(\xn; \param, \nnpar) = g(\xn; \nnpar)$. Additionally, we plot the CRB (under the PL model) for the sake of benchmarking.

Both scenarios involve a static jammer, transmitting with $\Ptx=10$\,dB.
We collected $10000$ observations over an area of $1$\,km$^2$, yielding, on average, 1 observation every $10 \times 10$\,m square. Since the signal power decreases very fast with respect to the distance from the source, weak power observations are not informative and buried in the noise floor. Only the $15$ most powerful observations were fed to the estimators. We repeated the data collection for different measurement noise variance and therefore different \gls{inr}, defined as $\inr = 10 \log_{10} \Ptx/\sigma^2\ .$
For each \gls{inr} level, $N_{\textrm{MC}} = 100$ Monte Carlo realizations were simulated to collect statistics about the estimators performances. To compare the different approaches we consider the \gls{rmse} computed over each dimension of $\param$ as
\begin{equation}
    \label{eq:rmse}
    \rmse_{\theta_i} = \sqrt{\frac{1}{N_{\textrm{MC}}} \sum_{n=1}^{N_{\textrm{MC}}} (\theta_i - \hat{\theta}_{i,n})^2}
\end{equation}

The NN model is a feed-forward network with 2 hidden layers of 200 and 100 neurons and a hyperbolic tangent activation function. The PL model in $(ii)$ and $(iv)$ is initialized with the same information available to the MLE, i.e. $\Ptx$ and $\gamma$. The models are trained over 200 epochs with a learning rate of $0.4$, exploiting the Adam optimizer~\cite{adam} and batch learning.
%
\vspace{-0.2cm}
\begin{figure}[htb]
  \centering
  \centerline{\includegraphics[width=8.5cm]{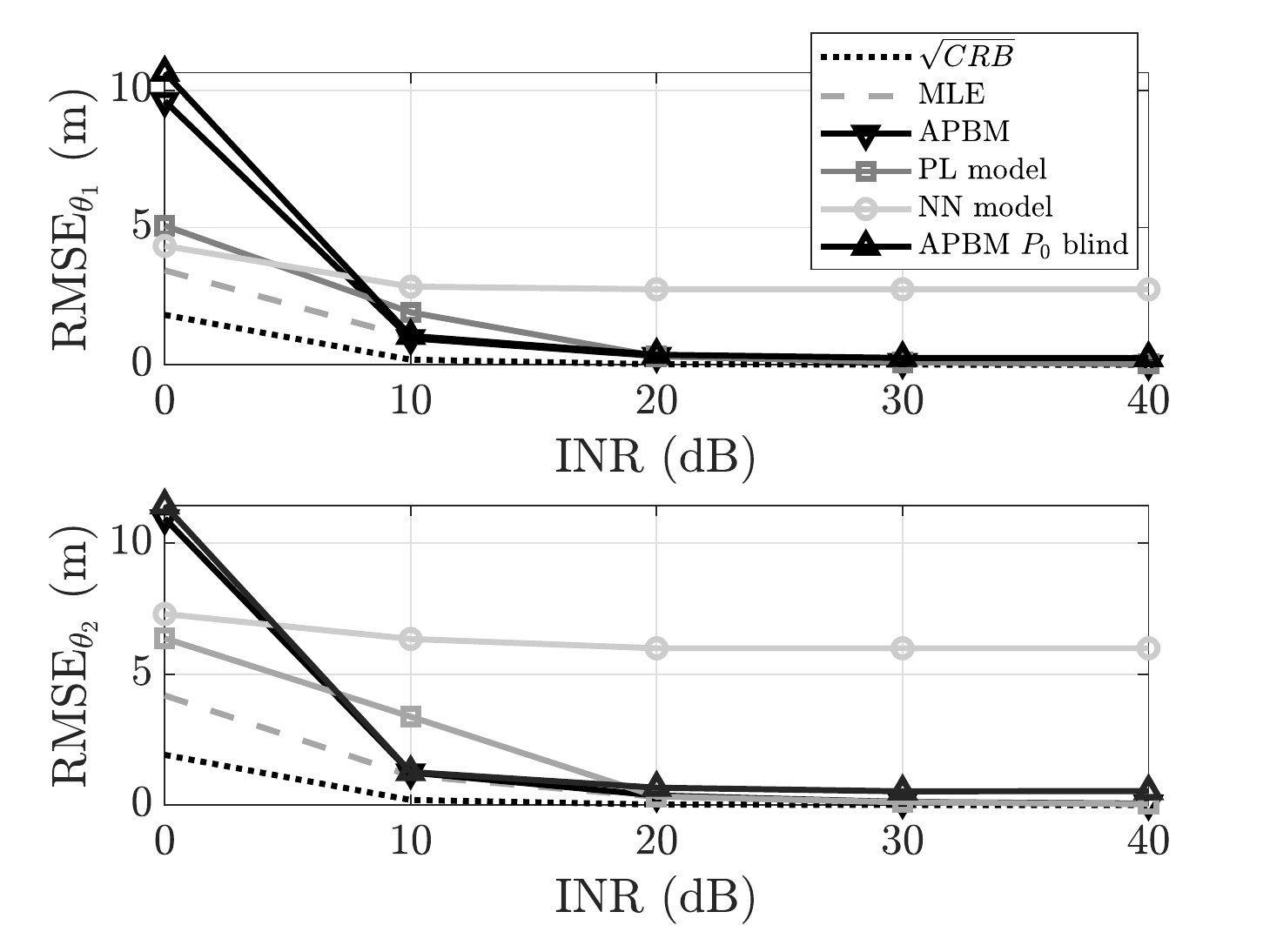}}
\vspace{-0.5cm}
\caption{RMSE in the PL propagation scenario.}
\vspace{-0.25cm}
\label{fig:PL_scenario}
\end{figure}

\begin{figure}[htb]
  \centering
  \centerline{\includegraphics[width=8.5cm]{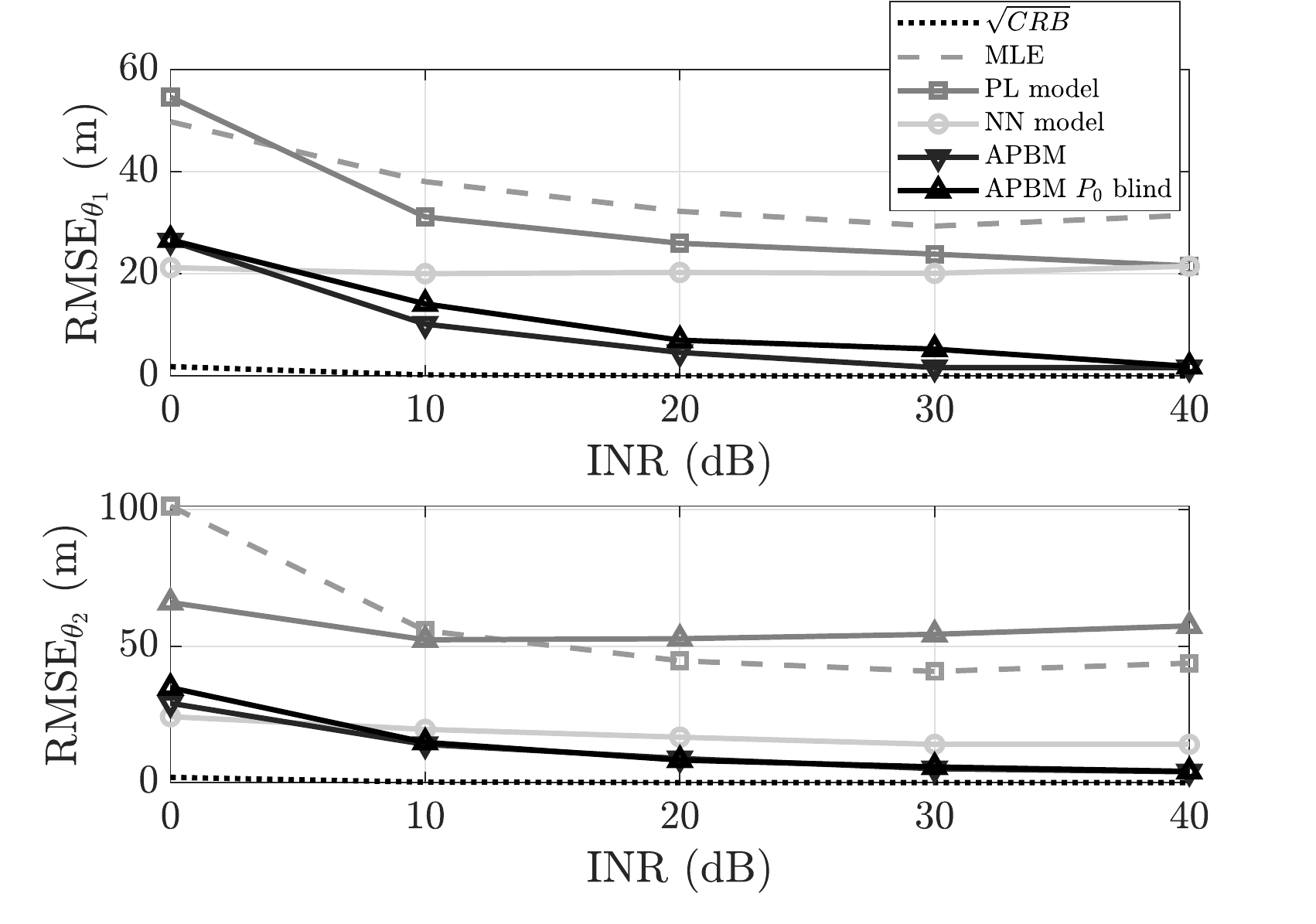}}
\vspace{-0.5cm}
\caption{RMSE in the ray tracing propagation scenario.}
\vspace{-0.5cm}
\label{fig:RT_scenario}
\end{figure}




In the \gls{pl} scenario, addressed in Figure~\ref{fig:PL_scenario}, the jamming signal propagates in nominal conditions, according to~\eqref{eq:rss_model}, and $(i)$ and $(iv)$ are based on a model that matches exactly this propagation rule. Not surprisingly, as the \gls{inr} increases, they are all approaching the CRB, which is also based on perfect modeling.
The use of the PL model component alone enables to closely follow the MLE, even at low INR values, where the fixed implementation settings might prevent such method to perform at its best. Indeed, it is worth stressing that the common settings described at the beginning of this section are kept constant for all the scenarios and methods. Moreover, the solution adopted to circumvent the pathloss formula singularities (see Section~\ref{sec:mle}) might affect the learning process in a systematic way for a given set of measurements and jammer locations. On the contrary, despite the simple propagation rule, the NN-only solution is unable to effectively learn the model and it is not able to improve further above 20\,dB of INR. When the separate components are joined again to realize the \gls{apbm} we observe a low RMSE, very close to the MLE, also around 10\,dB where the complete system $(ii)$ is outperforming the PL-only model $(iv)$. In these conditions, the NN component is probably helping the \gls{apbm} to overcome the learning flaws of the PL model. Conversely at very low INR the high amount of measurement noise has a counter-productive effect on the joint learning.
The \gls{apbm} can also work being $\Ptx$-blind $(iii)$, with no knowledge about the jammer's power and indeed converge to a correct estimation (Figure~\ref{fig:PL_scenario}). In this case the NN is effectively able to learn a constant term, which compensate for the missing jammer information, providing a fundamental aid.
%
The low RMSE values attained by the MLE and PL model in Figure~\ref{fig:PL_scenario} are possible thanks to a perfect knowledge of the propagation environment. When it comes to a realistic scenario such as the one in Figure~\ref{fig:RT_field}, few valid assumptions on the pathloss can be made. This results in the jammer location estimation performances shown in Figure~\ref{fig:RT_scenario}. The MLE and PL-only solution are far form the attainable minimum, instead  the \gls{apbm} and NN-only model are performing better. How close are their respective RMSE values strictly depends on how far the actual propagation model is from an ideal pathloss. The more mismatched is the PL model the closer the \gls{apbm} performance is to the NN, as in this case. It is worth noting that despite the large difference between the propagation scenarios, the APBM can seamlessly adapt to both without changes. Also, in this realistic scenario, the ability of the \gls{apbm} to perform a successful $\Ptx$-blind estimation is even more relevant, since this information is not usually available.




\section{Conclusion}
\label{sec:conclusion}
In this work we proposed a jamming source localization strategy using crowdsourced measurements. We investigated the use of an augmented physics-based model (APBM) approach that models the complex propagation channel with a physics-based pathloss model and a data-driven NN component, jointly estimating the position of the jamming source and the NN model parameters. 
The proposed method exhibited flexibility to characterize both ideal propagation conditions and complex urban scenarios, attaining a performance comparable to a perfect model-aware MLE in the first case, and better than a stand-alone NN in the latter. Moreover, such performance can be reached also with no information about the jamming source, thus without knowledge about the transmission power, a parameter not available in realistic scenarios. 
Further works will investigate the adaptability of the method in dynamic scenarios, involving a moving and intermittent jammer, considering also continual learning to address the changes in data distribution. To address privacy concerns, we intend to connect this work with federated learning schemes. 



\vfill
\clearpage

\bibliographystyle{IEEEtran}
\bibliography{bibfile,closas-ref}

\end{document}